\documentstyle[epsfig]{article}

\def\tou#1{{\lower1.2ex\hbox{$\longrightarrow$}\atop
        {\lower-.7ex\hbox{$\scriptscriptstyle #1 $}}}}
\def\lsim{{\lower1.2ex\hbox{$<$}\atop
        {\lower-.7ex\hbox{$\sim$}}}}
\def\gsim{{\lower1.2ex\hbox{$>$}\atop
        {\lower-.7ex\hbox{$\sim$}}}}

\def\be{\begin{equation}}
\def\ee{\end{equation}}

\begin{document}

\begin{titlepage}
\rightline {Si-95-10 \  \  \  \   }

\vspace*{2.truecm}

\centerline{\Large \bf  Determination of the Critical Point and Exponents }
\vskip 0.6truecm
\centerline{\Large \bf
        from Short-time Dynamics }
\vskip 0.6truecm

\vskip 2.0truecm
\centerline{\bf L. Sch\"ulke and B. Zheng}
\vskip 0.2truecm

\vskip 0.2truecm
\centerline{Universit\"at -- GH Siegen, D -- 57068 Siegen, Germany}

\vskip 2.5truecm

\abstract{The dynamic process for the two dimensional
three state Potts model in the critical domain
is simulated by the Monte Carlo method.
It is shown that the critical point can rigorously be located
from the universal short-time behaviour. This makes it possible
to investigate critical dynamics independently of
the equilibrium state. From the power law behaviour of
the magnetization the exponents $\beta / (\nu z)$ and $1/ (\nu z)$
are determined.

}

\end{titlepage}

\section{Introduction}

Recently more and more attention is drawn to the universal behaviour
of the critical dynamic process in the macroscopic short-time regime.
for the critical dynamic process of model A for the $O(N)$
vector model starting from a random state
with a small but non-zero initial
magnetization it has been argued with $\epsilon$-expansion up to two loop that
the magnetization undergoes a {\it critical initial increase}
\cite {jan89}.
Such a scenario has been simulated for the Ising model and the Potts model
by the Monte Carlo method \cite {li94,sch95},
and the new dynamic exponent $\theta$ which governs the initial
universal behaviour has directly been determined.
 The exponent $\theta$ may also be estimated indirectly from
the power law decay of the auto-correlation \cite {hus89,hum91},
and with the damage spreading technique \cite {gra95}.
Universality \cite {men94,oka95} as well as
a generalized finite size scaling \cite {li94,li95,li95a}
has also been confirmed.
Such an investigation of the short-time critical dynamics not only
enlarges our fundamental knowledge on critical phenomena but also
provides possible improvement on the numerical determination
of the dynamic and the static exponents.
The generalization and application to complex systems are also
undergoing \cite {hus89,lee95,rut95,blu92}.

Of special interest is here a previous paper by the authors where
the short-time dynamics for the
Potts model at the critical point was simulated
with the Monte Carlo Heat-bath
algorithm \cite {sch95}.
We have concentrated ourselves on
the critical initial increase of the magnetization
as well as the power law behaviour of the auto-correlation
and the second moment of the magnetization.
The new dynamic exponent $\theta$ has directly been measured and
the traditional
dynamic exponent $z$ as well as $\beta / \nu$ were also obtained.
The results relate the direct and indirect measurement
of $\theta$ to each other and provide confident support
for the dynamic scaling
in the short-time regime. Especially the measured
dynamic exponent $z$ is quite rigorous.
Compared with the distributed values of $z$ between
$z=2.2$ and $z=2.7$ from different simulations
\cite  {ayd85,tan87,bin81,ayd88},
the exponent $z$ we obtained supports the small one
\cite {ayd85,tan87}.

However some important problems remain open
in the numerical simulation of the short-time dynamics.
First, the critical point was {\it always input} in the numerical
investigation of the short-time dynamics.
 The critical point is usually
defined and measured in the equilibrium state of the systems..
It is be very interesting to know whether one can
define and locate the critical point from the short-time
dynamics itself or not. This would allow a complete investigation
of the short-time dynamics independent of the
equilibrium state. On the other hand, for the critical system
(with infinite volume) exactly
at the critical point the dynamic process will, in principle,
 never reach the equilibrium state
due to the critical slowing down, at least for the {\it local
dynamics}. Therefore it is also important from the practical point of view
that one can determine the critical point
from the dynamics itself.
Furthermore one may imagine that the critical point for the dynamic
system could be different from that for
the equilibrium state. This may conceptually broaden the scope of the
critical phenomena. Actually some discussion of this kind
has been made with respect to the damage spreading \cite {gra95a}.
One purpose of this letter is to clarify how to
locate rigorously the critical point from the short-time dynamics.

Secondly, how to determine the static exponent $\beta$ and $\nu$
from the short-time dynamics
in a rigorous way still requires further investigation.
It has been pointed out
that the determination of the static exponents from
the short-time dynamics is quite promising
since one does not enter the long-time regime
of the dynamic evolution where the
critical slowing down is severe \cite {li95,li95a}.
It has been demonstrated for the Ising model that the static exponents
$\beta$ and $\nu$ may be estimated from finite size scaling.
One can easily realize that the measurement of the static
exponents from the dynamics depends essentially on the quality of
the measurement of the dynamic exponent $z$.
In some cases a small relative error for $z$ may induce
big errors for the static exponents.
To determine the dynamic exponent $z$ independently with
the time-dependent Binder cumulant in principle works well.
But it happens sometimes that the cumulant is
somewhat fluctuating since it is constructed from
higher moments of the magnetization.

Therefore another purpose of this Letter is to provide an alternative way
to determine the static exponents $\beta$ and $\nu$
from the power law behaviour of the observables in the
beginning of the time evolution of the dynamic process.
 Compared with the method based on finite size scaling,
the measurement of the exponents from the power law behaviour
of the observables is more direct. The exponents may be extracted
from a single lattice and the finite size effect may easily be
under control.
Actually in reference \cite {sch95}
the static exponent $\beta/\nu$ has been obtained
from the power law increase of
 the second moment. However the result is still rough.
What one practically measures from the second moment of
the magnetization is $(d-2 \beta/\nu)/z$. As it was pointed out
\cite {sch95}, the value of
$\beta/\nu$ is much smaller than $d$ and $z$, and therefore
a small relative error for $z$ or $(d-2 \beta/\nu)/z$ will induce
a big error for $2 \beta/\nu$. We should search for a
better way.

\vspace{0.3cm}

Let us start with the determination of the critical point and
the exponent $1/(\nu z)$. Following the consideration
in \cite {sch95}, we first study the critical dynamic process
of model A for the
two dimensional three state Potts model
starting from a random initial state with small but
non-zero initial magnetization.
The Hamiltonian is given by
\be
{\cal H} = J \sum_{<i,j>} \delta_{\sigma_i\sigma_j};\qquad \sigma_i=-1,0,1.
\ee
According to the argument
of Janssen, Schaub and Schmittmann obtained with the renormalization
group method, one may expect a generalized scaling relation
\begin{equation}
M^{(k)}(t,\tau,m_{0})=b^{-k\beta/\nu}
M^{(k)}(b^{-z}t,b^{1/\nu}\tau,
b^{x_{0}}m_{0}).
\label{e1}
\end{equation}
to be valid also for the Potts model. for the magnetization ($k=1$),
taking $b=t^{1/z}$ and assuming $t^{x_{0}/z}m_{0}$
is small enough, we can expand with respect to $t^{x_{0}/z}m_{0}$
and obtain
\begin{equation}
M(t,\tau,m_{0})=m_0 t^{\theta}
F(t^{1/(\nu z)}\tau)
\label{e2}
\end{equation}
where $\theta = (x_0 - \beta /\nu)/z$.
It is clear that for $\tau=0$ we observe the so-called
 critical initial (power law) increase of the magnetization.
for $\tau \neq 0$, but still in the critical domain
the time evolution of the magnetization will be
modified by the scaling function $F(t^{1/(\nu z)}\tau)$.
Such a fact can be used for the determination
of the critical point.
In Fig. 1 the time evolution of the magnetization
for three different coupling $J=1.0025$, $J=1.0050$, and $J=1.0075$,
 with initial
magnetization $m_0=0.04$ and lattice size
$L=72$ are displayed with solid lines in double--log scale.
The coupling for the line in the centre, $J=1.0050$, corresponds
to the theoretical critical point for the infinite system.
The total number of samples with independent
initial configurations for each coupling $J$ is 600 000.
By a linear extrapolation in $J$ between the simulated curves
one can find a particular coupling
which gives the best power law behaviour for the magnetization.
This is the ``critical point'' for the finite system.
In principle there can also be some effect from the finite
initial magnetization
$m_0$. The dots in Fig. 1 represent the time evolution of the
magnetization at such a critical point $J_c=1.0055(8)$.
It is remarkable that the critical point located in this way is so
close to the exact one.
To estimate the critical point, we have linearly interpolated
the magnetization $M(t,J)$ between the measured
curves for $J=1.0025$ and $J=1.0075$. At the time $t$, the slope is
measured within a time interval $[t,t+15]$. Theoretically, at the
critical point $J_c$, the slope $\theta(t,J_c)$ is constant with respect
to $t$. Practically
the critical point is determined by searching for the minimum
square deviation of the slope $\theta(t,J)$ from a constant.
In Fig.~2 this deviation
is shown
when the coupling is varying around the critical value.
The clear minimum is a confident indication for the
critical point. It has been pointed out that
in the measurement of the exponent $\theta$, the microscopic time scale
$t_{mic}$ for the Heat-Bath algorithm is ignorably small.
Therefore in the above estimate of the critical point,
the time interval is taken to be $[1,100]$.
Actually the time interval may be taken even smaller, e.g. $[1,50]$.
This is really the physics from the {\it short-time
regime} of the dynamic evolution.
Furthermore, examining the result for smaller
lattice size $L=32$ and somewhat bigger initial magnetization
$m_0=0.08$, one can realize that the effects of finite size and $m_0$
for $L=72$ and $m_0=0.04$ are already very small, at least
in the determination
of the critical point. This is an advantage for determining
the critical point from the short-time dynamics.

{\it Theoretically} the critical point could be located from
short-time behaviour of any observable in the critical domain.
However practically many of them are not very sensitive
to the coupling. Also to determine the critical point
efficiently the dependence of the chosen observable on
the critical exponents should be as small as possible.
For example, the second moment of the magnetization and the
auto-correlation
are not good choices for the determination of
the critical point. In reference \cite {li94} it was
mentioned that one might be able to
 locate the critical point from the time-dependent
cumulant. But this also appears to have some complication since higher
moments are involved.

To determine the exponent $1/(\nu z)$, we differentiate
$\ln \ M(t,\tau,m_{0})$ with respect to $\tau$ at the critical point,
and for small enough $t^{x_{0}/z}m_{0}$
\begin{equation}
\partial_\tau \ \ln \ M(t,\tau,m_{0})|_{\tau=0}= t^{1/(\nu z)} \
\partial_\tau \ \ln \ F(\tau')|_{\tau'=0}
\label{e3}
\end{equation}
Therefore
$\partial_\tau \ln \ M(t,\tau,m_{0})|_{\tau=0}$ also shows a power law
increase.
In the numerical simulation
we have approximated the differentiation by the
difference between $J=1.0025$ and $J=1.0075$.
This derivative is plotted in Fig. 3 in double--log scale.
{}From the slope one can estimate the critical exponent $1/(\nu z)$.
Using the exponent measured from the auto-correlation as input,
one obtains the static exponent $1/\nu$.
 The result together with the exponent $z$ and the measured
critical point is given in table 1.
Due to some possible effect of $t_{mic}$, we have made the measurement
within the time interval $[20,100]$.
Compared with the exact
value $1/\nu=1.2$ the measured value is slightly bigger.
This comes probably from the approximation of the differentiation
by the difference. To improve the quality of $1/\nu$
we might represent alternatively
the derivative by higher moments \cite {fer91}.
Here we should point out that the slight difference between
the values for $z$ in table 1
and that in reference \cite {sch95} comes from some more serious
consideration of the effect of $t_{mic}$ and the finite lattice size
\cite{oka95}.

\begin{table}[h]\centering
$$
\begin{array}{c|c|c|c}
      J_c     &       z     &    \beta/\nu &  1/\nu \\
\hline
   1.0055(8) & 2.196(8) & 0.1330(3)  &   1.24(3)\\
\end{array}
$$
\caption{\footnotesize
Results for $J_c$, $\beta/\nu$ and $1/\nu$ obtained in this paper.
The value for $z$ used as an input has been taken from the previous
investigation.
}
\label{t1}
\end{table}

Now we turn to the determination of the static exponent
$\beta/(\nu z)$. It has been discussed before that it is not the best way
to measure it from the second moment. Due to the very short
spatial correlation length in the very beginning of the time evolution
of the dynamic process the second moment
behaves as $M^{(2)}(t,L) \sim L^{-d}$. Therefore
obtaining a good power law behaviour for the second moment
is not so easy, especially for
relative big lattices in the short-time regime.
 Fortunately recent investigations of the dynamic
Ising model \cite {sta92,li95a}
shows that from the power law {\it decay}
of the magnetization
\be
M(t) \sim t^{-\frac{\beta}{\nu z}}
\ee
in the dynamic process starting from
{\it a completely ordered state}, i.e. $m_0=1$,  it is possible to
 estimate the
exponent $\beta/(\nu z)$. With $z$ in hand, we can obtain
the static exponent $\beta/\nu$. Here we have simulated
such a dynamic process for the Potts model.
We have pointed out before that the critical point we estimated is very
close to the exact one. Therefore in the simulation here we take
the exact value for the critical point for simplicity.
The results support that universality and scaling
appear to be valid also in a quite early stage
of the time evolution and one can really obtain
the exponent $\beta/(\nu z)$ quite accurately in the short--time dynamics.
In Fig. 4 the time evolution of the magnetization
is displayed for lattice sizes $L=72$ and $L=144$ in double--log scale.
The total number of samples for independent initial
configurations is $200 000$ for $L=72$ and $80 000$ for $L=144$.
One can hardly see any difference between both curves, i.e., there
is already an invisible finite size effect for lattice size $L=72$.
{}From the slope of the curves
one can estimate the exponent $\beta/(\nu z)$.
   Detailed analysis reveals that in this case
$t_{mic} \sim 50$, which
is bigger than that in the measurement of $\theta$.
However the behaviour becomes very stable after $t_{mic}$.
Measuring the slope in a time interval
$[100,250]$ and using the value of $z$ as input,
one obtains the static exponent $\beta/\nu=0.1326(6)$
for $L=72$ and $\beta/\nu=0.1333(2)$ for $L=144$.
The errors
are estimated by dividing the samples into two or six groups.
The average value is given in table 1.
The static exponent $\beta/\nu$ obtained in this way is very rigorous.
In general the quality of the measurement for the first moment
is better than that of the second moment.
Besides the magnetization in this process is strongly self-averaging
and therefore one can easily simulate bigger
lattices. Furthermore the error of the input $z$ will
now have almost no effect on the quality of the exponent $\beta/\nu$
since the value of $\beta/(\nu z)$ is much smaller than
that of $z$. This situation is very different from
that for the measurement of the second moment.

In conclusion, we have numerically
simulated the critical dynamic process of model A
for the two dimensional three state Potts model starting
from both a random state with
small initial magnetization and a completely ordered state.
{}From the universal power law behaviour of
the magnetization in the short-time regime of the dynamic
evolution, the critical point and the static exponents
are obtained. The results provide further confirmation
for the dynamic scaling in the short-time dynamics,
including the extension to the dynamic process starting from a completely
ordered state. Especially the measurement of the
critical point and the exponent $\beta/\nu$ is very rigorous.
This makes it possible to study the short-time dynamics
completely independent of the equilibrium state.
The determination of the static properties from
the short-time dynamics is not only conceptually interesting
but also practically applicable.
The application of this technique to the unsolved critical
systems is being carried out.

\begin{figure}[p]\centering
\epsfysize=12cm
\epsfclipoff
\fboxsep=0pt
\setlength{\unitlength}{1cm}
\begin{picture}(13.6,12)(0,0)
\put(0,0){{\epsffile{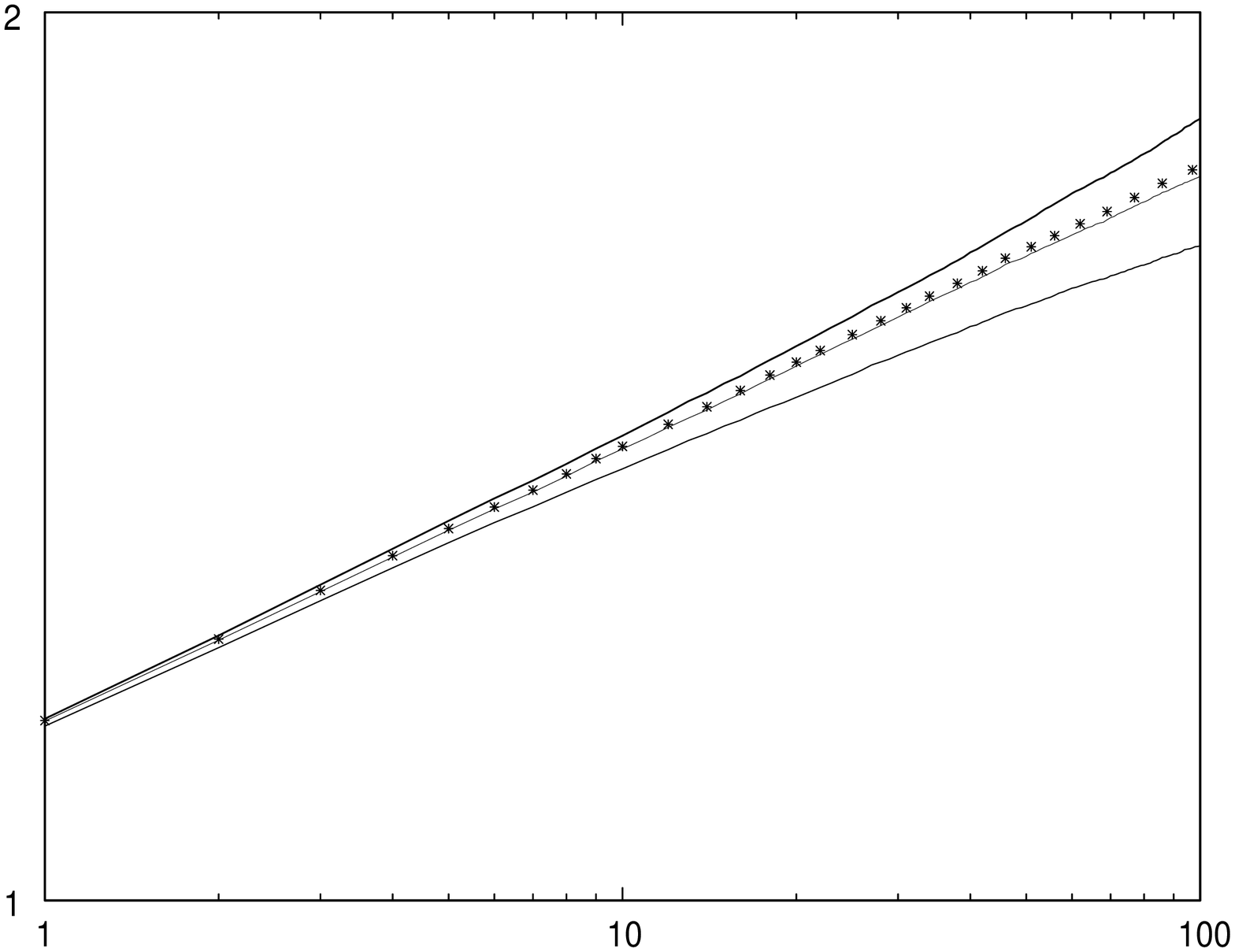}}}
\put(11.5,7.9){\footnotesize J=1.0025}
\put(11.5,7.4){\footnotesize J=1.0055(8)}
\put(11.5,7.1){\footnotesize J=1.0050}
\put(11.5,6.5){\footnotesize J=1.0075}
\end{picture}
\caption{\footnotesize
The magnetization $M(t)$ for initial magnetization $m_0=0.04$
plotted versus time in a double-log scale. The solid lines correspond to
three different values for $J$, $J=1.0025$, $J=1.0050$ and $J=1.0075$.
The dotted curve represents the magnetization
for the critical point $J_c=1.0055$ which shows the best power law
behaviour.
}
\label{f1}
\end{figure}

\begin{figure}[p]\centering
\epsfysize=12cm
\epsfclipoff
\fboxsep=0pt
\setlength{\unitlength}{1cm}
\begin{picture}(13.6,12)(0,0)
\put(0,0){{\epsffile{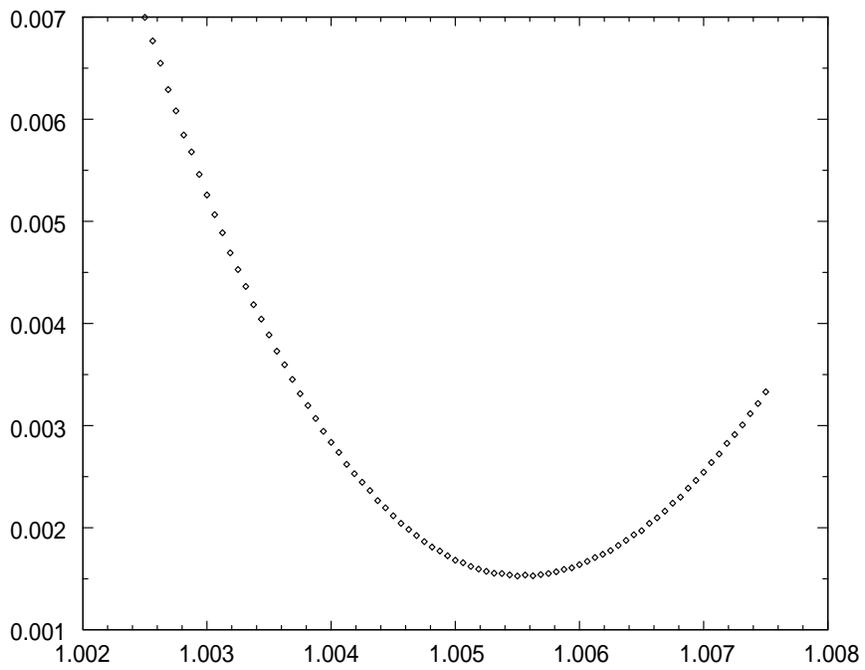}}}
\end{picture}
\caption{\footnotesize
Deviation from the power law behaviour when $J$ is varied
around the critical point.
}
\label{f2}
\end{figure}

\begin{figure}[p]\centering
\epsfysize=12cm
\epsfclipoff
\fboxsep=0pt
\setlength{\unitlength}{1cm}
\begin{picture}(13.6,12)(0,0)
\put(0,0){{\epsffile{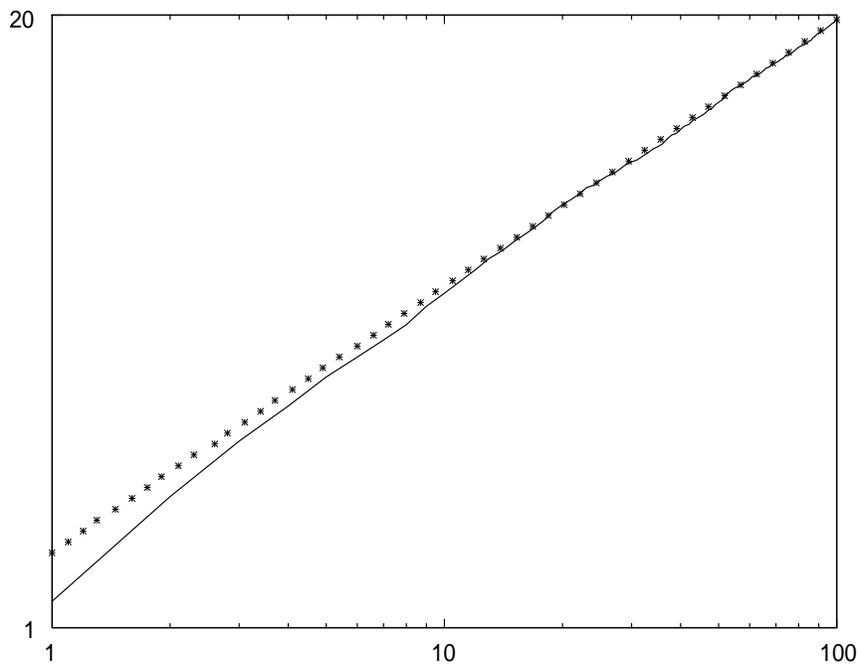}}}
\end{picture}
\caption{ \footnotesize
The derivative $\partial_\tau\ln M(t,\tau,m_0)|_{\tau=0}$
plotted versus time in a double-log scale. In order to avoid possible
effects of $t_{mic}$, the slope has been calculated from the time
interval $t=[20,100]$. The dotted curve represents a straight line with
the measured slope.
}
\label{f3}
\end{figure}

\begin{figure}[p]\centering
\epsfysize=12cm
\epsfclipoff
\fboxsep=0pt
\setlength{\unitlength}{1cm}
\begin{picture}(13.6,12)(0,0)
\put(0,0){{\epsffile{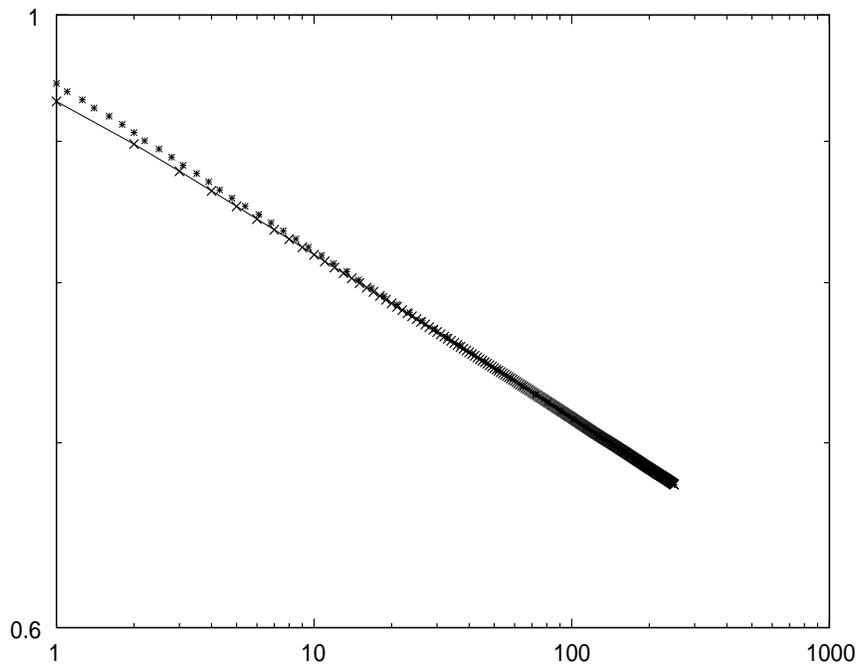}}}
\end{picture}
\caption{\footnotesize
The magnetization $M(t)$ versus time in a double-log scale. The
solid line and the crosses on that line represent the results for
$L=72$ and $L=144$. The slope has been obtained from the time interval
$t=[100,250]$ in order to avoid effects from $t_{mic}$. The dotted
curve is the straight line with the measured slope.
}
\label{f4}
\end{figure}

\end{document}